\documentstyle[12pt]{article}
\begin{document}
\begin{titlepage}
\scrollmode
\begin{center}
{\Large\bf On a Generalized Oscillator~:}

\vspace{0.5cm}
{\Large\bf Invariance Algebra and Interbasis Expansions}
\end{center}

\vspace{0.5cm}
\begin{center}
{\bf Y.M.~HAKOBYAN,$^1$
         M.~KIBLER,$^2$
     G.S.~POGOSYAN,$^1$
and  A.N.~SISSAKIAN$^1$}
\end{center}

\begin{center}
{$^1$Bogoliubov Laboratory of Theoretical Physics,}
{Joint Institute for Nuclear}\\
{Research, 141980 Dubna, Moscow region, Russia}
\end{center}

\begin{center}
{$^2$Institut de Physique Nucl\'eaire de Lyon,}
{IN2P3-CNRS et Universit\'e Claude}\\
{Bernard, 43 Bd du 11 November 1918,}
{F-69622 Villeurbanne Cedex, France}
\end{center}

\vspace{2cm}
\begin{abstract}

This article deals with a quantum-mechanical system which 
generalizes the ordinary isotropic harmonic oscillator system. We 
give the coefficients connecting the polar and Cartesian bases for 
$D=2$ and the coefficients connecting  the  Cartesian  and  cylindrical 
bases as well as the  cylindrical  and  spherical  bases for $D=3$. 
These interbasis expansion coefficients are found to be analytic
continuations to real values of their arguments of the
Clebsch-Gordan coefficients for the group SU(2).
For $D=2$, the superintegrable character for the generalized
oscillator system is investigated from the points of view of
a quadratic invariance algebra. 

\end{abstract}

\vskip 6.5 true cm
\noindent 
Paper to appear in the Proceedings of the {\bf VIII International Conference on
Symmetry Methods in Physics} (JINR, Dubna, Russia, 28 July -- 2 August 1997). 
The Proceedings of the Conference will be published in the Russian Journal of
Nuclear Physics (Yadernaya Fizika). 

\vfill
\thispagestyle{empty}
\end{titlepage}

\newpage


\section{Introduction}

During the last 30 years, superintegrable dynamical systems have been the
object of considerable interest (see [1-10] and references therein). 
In particular, numerous works have been devoted to the search for dynamical
invariance algebras (especially quadratic algebras) of nonrelativistic systems
with potentials presenting singularities. Such systems are important in various
fields (e.g., Aharonov-Bohm effect, Dirac or Schwinger monopoles, confining
problems, supersymmetry, etc.). 

It is the aim of this paper to investigate the system with the potential 
\begin{eqnarray}
\label{I1}
V   = \sum_{a=1}^{D} V_a, \quad 
V_a = \frac{1}{2} \Omega^2 x_a^2 
    + \frac{1}{2}  P  \frac{ 1 }{x_a^2}, \quad P=k_a^2 - \frac{1}{4} 
\end{eqnarray}
where $\Omega > 0$ and $k_a^2 > 0$ ($a = 1, 2, \cdots, D$). This system was
already discussed for $D=2$ by the late Professor Smorodinsky and his 
collaborators~\cite{SMORO} from a classical and quantum-mechanical point of
view. We shall  be concerned here mainly  with $D=2$ and 3 for
which the spectrum of the Schr\"odinger equation
\begin{eqnarray} 
\label{SCH}
H \Psi = E \Psi, \quad H = - \frac{1}{2} \Delta + V 
\end{eqnarray} 
shall be given. Emphasis shall be put 
on interbasis expansions in terms of
analytic continuation of  Clebsch-Gordan coefficients (CGc's)  for the group 
SU(2). As another important result, we shall introduce a quadratic invariance
algebra in the $D=2$ case. 

\section{$D$-dimensional case}

We briefly consider here the $D$-dimensional case in Cartesian 
coordinates. We start with $D=1$ and look for a solution of the  
one-dimensional equation (\ref{SCH}) for the potential $V_1$, 
see (\ref{I1}), with $x_1 \equiv x$ and $k_1 \equiv k$. 
The resolution of this equation, with the conditions $\Psi(x)
\rightarrow 0$ as $x \rightarrow 0$ and $\infty$, leads to the 
normalized wavefunction 
\begin{eqnarray}
\label{C4}
\Psi_{n}(x; \pm k) =
 \sqrt\frac{ {\Omega}^{\frac{1}{2}} n! } {\Gamma(n \pm k +1)}
\left( \sqrt{\Omega x^2} \right) ^{\frac{1}{2} \pm k}
{\rm exp}\left(- \frac{\Omega}{2} x^2 \right)
L_{n}^{\pm k}(\Omega{x}^2),   \quad   n \in {\bf N} 
\end{eqnarray}
where $L_{n}^{\nu}$ is an associated Laguerre polynomial~\cite{E1}.
The normalization is such that
\begin{eqnarray}
2 \int_{0}^{\infty}
\Psi_{n'}(x; \pm k)^*
\Psi_{n }(x; \pm k) d x = \delta_{n' n}
\end{eqnarray}
The discrete energy spectrum is given by 
$$ 
E = \Omega (2n \pm k + 1)
$$ 
Only the sign $+$ may be taken in front of $k$ when $k > \frac{1}{2}$. For 
$0 < k < \frac{1}{2}$, both the signs $+$ and $-$ are admissible. 
For $k = (\frac{1}{2})^{-}$, due to the connecting
formulas~\cite{ERD} 
between the (even and odd) Hermite polynomials ${\cal H}_p (x)$ and
the Laguerre polynomials $L_n^{(\pm \frac{1}{2})}(x^2)$ and by putting
$p=2n+1$  for  the sign $+$ and  $p=2n$  for the sign $-$, we immediately 
have 
$$ 
\Psi_{n}\left(x; \pm \frac{1}{2}\right) =
\left(\frac{\Omega}{\pi}\right)^{\frac{1}{4}}
\frac{1} { \sqrt{ 2^{p} p! } } {\rm exp}\left(- \frac{\Omega}{2} x^2 \right)
{\cal H}_{p}(\sqrt{\Omega x^2})
$$ 

We now deal with the $D$-dimensional case. In this case, the Cartesian 
wavefunction, that vanishes when $x_a \rightarrow 0$ and $\infty$ 
($a=1, 2, \cdots, D$), is 
$$ 
\Psi_{\bf n}( {\bf x} ; {\bf k} ) = \prod^{D}_{a=1} \Psi_{n_a}(x_a ; \pm k_a) 
$$ 
where ${\bf n} =     n_1  \cdots      n_{D}$ with $n_a \in {\bf N}$, 
      ${\bf x} =     x_1, \cdots,     x_{D}$ and 
      ${\bf k} = \pm k_1, \cdots, \pm k_{D}$. The energy is
$$ 
E = \Omega \left[ 2n + D + \sum_{a=1}^D (\pm k_a) \right]
$$ 
where $n= n_1+n_2+ \cdots +n_D$ is the principal quantum number.

\section{Two-dimensional case}

\subsection{Cartesian basis}

In Cartesian coordinates ($x_1 \equiv x$, $x_2 \equiv y$), the wavefunction is
 \begin{eqnarray}
 \label{P1}
 \Psi_{n_1n_2}(x,y ; \pm k_1, \pm k_2) = \Psi_{n_1}(x ; \pm k_1)
                                         \Psi_{n_2}(y ; \pm k_2)
 \end{eqnarray}
where $\Psi_{n_a}$ (with $a = 1,2$) are given by (\ref{C4}). Note that 
we have the new constant of motion 
\begin{eqnarray}
\label{P200}
N = \frac{1}{4\Omega} \left( D_{xx} - D_{yy} +
 \frac{k_1^2 - \frac{1}{4}}{x^2}
-\frac{k_2^2 - \frac{1}{4}}{y^2} \right)
\end{eqnarray}
(in addition to the energy), where 
$D_{\alpha \beta}= - \partial_{\alpha \beta} + \Omega^2 {\alpha} {\beta}$ 
is the Demkov tensor~\cite{DEM}. 

\subsection{Polar basis}

In polar coordinates ($\rho, \varphi$), the potential (\ref{I1}) reads
$$ 
V=\frac{1}{2} \Omega^2 \rho^2 +
\frac{1}{2 \rho^2}\left(
\frac{k_1^2 - \frac{1}{4}}{\cos^2\varphi} +
\frac{k_2^2 - \frac{1}{4}}{\sin^2\varphi}
\right)
$$ 
for which Eq.~(\ref{SCH}) may be separated by seeking 
a solution in the form $R (\rho) \Phi (\varphi)$. 
This leads to the system of coupled differential equations 
\begin{eqnarray}
\label{P67}
\left( d_{\varphi \varphi} 
+ A^2 - \frac{k_1^2 - \frac{1}{4}}{\cos^2\varphi}
-       \frac{k_2^2 - \frac{1}{4}}{\sin^2\varphi} \right) \Phi
= 0, \quad \left[ 
\frac{1}{\rho} d_{\rho} \left( \rho d_{\rho}   \right)
+       2E- \Omega^2\rho^2 -\frac{A^2}{\rho^2} \right] R 
= 0
\end{eqnarray}
where $A$ is a polar separation constant.

 The solution
 $\Phi(\varphi) \equiv \Phi_{m}(\varphi; \pm k_1,\pm k_2)$
 of the angular equation in (\ref{P67}) 
 with the conditions
\begin{equation}
\label{P8}
\Phi(0) = \Phi( \frac{\pi}{2} ) = 0
\end{equation}
is easily found to be
\begin{eqnarray}
\label{P9}
\Phi(\varphi) 
& = & \sqrt\frac{(2m \pm k_1 \pm k_2+1)m!
\Gamma(m \pm k_1\pm k_2+1)}
{2 \Gamma(m \pm k_1 +1)
\Gamma(m \pm k_2+1)} \nonumber \\ 
& \times &  
(\cos\varphi)^{\frac{1}{2} \pm k_1}
(\sin\varphi)^{\frac{1}{2} \pm k_2} P_m^{(\pm k_2, \pm k_1)}(\cos2\varphi) 
\end{eqnarray}
where $m \in {\bf N}$ and $P_{n}^{(\alpha,\beta)}$ denotes a Jacobi polynomial.
The normalization is such that
\begin{eqnarray}
\label{aadd}
4 \int_{0}^{\frac{\pi}{2}}
  \Phi_{m'}(\varphi; \pm k_1,\pm k_2)^*
  \Phi_{m }(\varphi; \pm k_1,\pm k_2)   d{\varphi} =
  \delta_{m'm}
\end{eqnarray}
Then, the separation constant $A$ is quantized as
\begin{equation}
\label{P10}
A = 2m \pm k_1 \pm k_2 + 1
\end{equation}

The radial solution $R(\rho) \equiv R_{n_\rho m}(\rho; \pm k_1, \pm k_2)$ 
in (\ref{P67}) is
\begin{equation}
\label{P11}
R(\rho) = 
\sqrt{\frac{2\Omega n_\rho!}
{\Gamma(n_\rho + 2m \pm  k_1 \pm k_2 +2)}}
\left( \sqrt{\Omega \rho^2} \right)^{A}
{\rm exp}\left(- \frac{\Omega}{2}\rho^2\right)
L_{n_\rho}^A (\Omega \rho^2) 
\end{equation}
where $n_\rho \in {\bf N}$ is the radial quantum number.
The function $R$ satisfies the orthogonality relation
$$ 
\int_{0}^{\infty}
R_{n_\rho^{'} m}(\rho; \pm k_1, \pm k_2)
R_{n_\rho     m}(\rho; \pm k_1, \pm k_2) \rho d\rho =
\delta_{n_\rho^{'} n_\rho}
$$ 

The energy $E$ corresponding to the $n + 1$ wavefunctions 
$$
\Psi_{n_{\rho}m}(\rho, \varphi ; \pm k_1, \pm k_2) 
\equiv R (\rho) \Phi (\varphi)
$$ 
(with $n = n_{\rho} + m$ fixed) is 
\begin{equation}
\label{P12}
E = \Omega (2n  \pm k_1 \pm k_2 + 2), \quad n \in {\bf N} 
\end{equation}
where $n$ is the principal quantum number. 
Note that only the sign $+$ in front of $k_{1}$ and $k_{2}$
has to be taken when $k_{1} > \frac{1}{2}$ and $k_{2} > \frac{1}{2}$.
In the case $0 < k_{a} < \frac{1}{2}$ 
(with $a=1,2$), Eq.~(\ref{P9}) shows that 
for each $n$ we have four levels corresponding to 
($\pm k_{1}, \pm k_{2}$). The degeneracy of the level with the principal 
quantum number  $n$ is $n+1$. This degeneracy is identical to the one 
of the isotropic oscillator in two dimensions, for which the degeneracy 
group is SU(2). 

For $k_{1} = k_2 = (\frac{1}{2})^{-}$, we have 
$A(  \frac{1}{2},   \frac{1}{2}) = 2m +2$, 
$A(- \frac{1}{2}, - \frac{1}{2}) = 2m   $ and  
$A(  \frac{1}{2}, - \frac{1}{2}) = 
 A(- \frac{1}{2},   \frac{1}{2}) = 2m +1$. 
Then, by using the connecting formulas~\cite{ERD} between Jacobi and 
Chebychev polynomials, we obtain the four following wavefunctions~\cite{MPS} 
\begin{eqnarray}
\label{P17}
\Psi_{2n, 2m}(\rho, \varphi)     &=& \frac{1}{\sqrt\pi}
R_{2n,2m}(\rho)     \cos2m    \varphi, 
\qquad {\tilde n} = 2n
\\[2mm]
\label{P18}
\Psi_{2n+2, 2m+2}(\rho, \varphi) &=& \frac{1}{\sqrt\pi}
R_{2n+2,2m+2}(\rho) \sin(2m+2)\varphi,
\qquad {\tilde n} = 2n+2
\\[2mm]
\label{P19}
\Psi_{2n+1, 2m+1}(\rho, \varphi) &=& \frac{1}{\sqrt\pi}
R_{2n+1,2m+1}(\rho) \cos(2m+1)\varphi,
\qquad {\tilde n} = 2n+1
\\[2mm]
\label{P20}
\Psi_{2n+1, 2m+1}(\rho, \varphi) &=& \frac{1}{\sqrt\pi}
R_{2n+1,2m+1}(\rho) \sin(2m+1)\varphi,
\qquad {\tilde n} = 2n+1
\end{eqnarray}
corresponding to the energy $E = \Omega({\tilde n}+1)$. 
In Eqs.~(\ref{P17})-(\ref{P20}), we have 
$$ 
R_{p,t}(\rho) = \sqrt{\frac{2\Omega(\frac{p - t}{2})!}
{(\frac{p+t}{2})!}}
\left( {\sqrt{\Omega \rho^2}} \right)^{t}
{\rm exp}\left(-\frac{\Omega}{2}\rho^2\right)
L_{\frac{p-t}{2}}^{t}(\Omega{\rho^2})
$$ 
to be compared with the corresponding result for the
ordinary circular oscillator.

To close this section, let us mention that 
\begin{eqnarray}
\label{P22}
M = \frac{1}{4} \left( - \partial_{\varphi \varphi} + 
\frac{k_1^2 - \frac{1}{4}}{\cos^2\varphi} +
\frac{k_2^2 - \frac{1}{4}}{\sin^2\varphi} \right)
= \frac{1}{4} \left[  L_z^2 + (x^2 + y^2) \left( 
\frac{k_1^2 - \frac{1}{4}}{x^2} +
\frac{k_2^2 - \frac{1}{4}}{y^2}           \right)
              \right]
\end{eqnarray}
is a polar constant of motion, the eigenvalues of which 
are $A$ (see (\ref{P10})). 

\subsection{Connecting Cartesian and polar bases}

According to first principles in quantum mechanics, we have
\begin{equation}
\label{CP1}
{\Psi}_{n_1  n_2} = \sum_{m=0}^n W_{n_1 n_2}^m (\pm k_1, \pm k_2)
{\Psi}_{n_\rho m} 
\end{equation}
where $n_\rho + m = n_1 + n_2 = n$. In Eq.~(\ref{CP1}), it is
understood that the wavefunctions both in the left-
and right-hand sides are written in  polar
coordinates $(\rho,\varphi)$. Furthermore, by using the asymptotic 
formula for the associated Laguerre polynomials, 
Eq.~(\ref{CP1}) yields an equation that depends only on the variable
$\varphi$.
Thus, by using the orthonormality property of the function $\Phi$
with respect to the quantum number $m$, we obtain 
\begin{eqnarray}
\label{CP4}
           W_{n_1 n_2}^m   (\pm k_1, \pm k_2)
= (-1)^{m} B_{n_1 n_2}^m   (\pm k_1, \pm k_2) 
           E_{n_1 n_2}^{m} (\pm k_1, \pm k_2) 
\end{eqnarray}
where
\begin{eqnarray}
\label{CP5}
E_{n_1 n_2}^{m}(\pm k_1, \pm k_2)
=
2 \int_{0}^{\frac{\pi}{2}}(\sin\varphi)^{2n_2 + 1 \pm 2k_2}
(\cos\varphi)^{2n_1+1\pm 2k_1}\, P_{m}^{(\pm k_2, \pm k_1)}
(\cos 2 \varphi) d \varphi
\end{eqnarray}
and 
$$
B_{n_1 n_2}^m (\pm k_1, \pm k_2) = \sqrt{2m \pm k_1 \pm k_2 + 1}
$$
\begin{eqnarray}
\label{CP6}
\times \sqrt{
\frac{(n-m)!m!\Gamma(m \pm k_1 \pm k_2+1)
\Gamma(n + m \pm k_1 \pm k_2 + 2)}
{n_1!n_2!\Gamma(m \pm k_1+1)\Gamma(m\pm k_2+1)
\Gamma(n_1 \pm k_1+1)\Gamma(n_2 \pm k_2+1)}}
\end{eqnarray}
By making the change of variable $x=\cos 2\varphi$ and by
using the Rodrigues formula~\cite{ERD} for the Jacobi polynomial, 
Eqs.~(\ref{CP4})-(\ref{CP6}) lead to the integral representation
$$ 
W_{n_1 n_2}^m(\pm k_1, \pm k_2) 
$$
$$
=
\sqrt{\frac{(2m\pm k_1 \pm k_2+1)(n-m)!\Gamma(m \pm k_1 \pm k_2+1)
\Gamma(n+m\pm k_1 \pm k_2+2)}{n_1!n_2!m!\Gamma(m\pm k_1+1)
\Gamma(m\pm k_2+2)\Gamma(n_1 \pm k_1+1)\Gamma(n_2 \pm k_2+1)}}
$$ 
\begin{eqnarray}
\label{CP8}
\times \frac{1}{2^{n_1+n_2+m\pm k_1 \pm k_2+1}}
\int_{-1}^{1}(1-x)^{n_2}(1+x)^{n_1}\frac{d^m}{dx^m}
[(1-x)^{m\pm k_2}(1+x)^{m \pm k_1}]dx
\end{eqnarray}
for the interbasis expansion coefficients $W_{n_1 n_2}^m(\pm k_1, \pm k_2)$.

Equation (\ref{CP8}) can be compared with the integral 
representation~\cite{VMK} for the CGc's 
$\langle ab {\alpha}{\beta} | c \gamma \rangle$
of the group SU(2). This yields
\begin{equation}
\label{CP10}
W_{n_1 n_2}^m(\pm k_1, \pm k_2)=(-1)^{n_1-n_\rho}
 \langle ab {\alpha}{\beta} | c \gamma \rangle 
\end{equation}
with  $2 a      = {n_1+n_2 \pm k_1}      $, 
      $2 b      = {n_1+n_2\pm k_2}       $,
      $2 c      = 2 m     \pm k_1 \pm k_2$, 
      $2 \alpha = {n_1-n_2 \pm k_1}      $  and  
      $2 \beta  = {n_2-n_1\pm k_2}       $. 
Since the quantum numbers in (\ref{CP10}) are not
necessarily integers or half of odd integers, 
the coefficients for the expansion of the Cartesian
basis in terms of the polar basis may be considered as analytical
continuation of the SU(2) CGc's. 

The inverse of Eq.~(\ref{CP1}), viz., 
\begin{equation}
\label{CP11}
{\Psi}_{n_\rho m} = 
\sum_{n_1=0}^n 
{\tilde W}_{n_\rho m}^{n_1}(\pm k_1, \pm k_2)
{\Psi}_{n_1n_2} 
\end{equation}
follows from the orthonormality property of the SU(2) CGc's. Thus, the relation 
$$ 
{\tilde W}_{n_\rho   m}^{n_1} (\pm k_1, \pm k_2) =
        W _{n_1    n_2}^m     (\pm k_1, \pm k_2)
$$ 
gives the expansion coefficients in (\ref{CP11}). The SU(2) CGc's can be 
expressed~\cite{VMK} 
in terms of the hypergeometric function $_3F_2(1)$, so that 
Eq.~(\ref{CP10}) can be rewritten as
$$
W_{n_1 n_2}^m (\pm k_1, \pm k_2) = 
(-1)^{n_2}
\frac{n!\Gamma(n_2+m\pm k_2+1)}
{\sqrt{n_\rho!\Gamma(m\pm k_1+1)\Gamma(m\pm k_2+1)}} 
$$
$$
\times \sqrt{ (2 m \pm k_1 \pm k_2 + 1)
\frac{\Gamma(n_1\pm k_1+1)\Gamma(m\pm k_1\pm k_2+1)}
{n_1! n_2! m!
\Gamma(n_2\pm k_2+1)\Gamma(n+m\pm k_1\pm k_2+2)}}   
$$
$$
\times{_3F_2}\biggl(\matrix{-n-m\mp k_1\mp k_2-1, \mbox{  } -n_2, 
\mbox{  } -m\hfill\cr
-n_1-n_2, \mbox{  } -n_2-m \mp k_2\hfill\cr}
\bigg\vert 1 \biggr)                                
$$
By using symmetry properties for $_3F_2(1)$, we arrive at the expression 
$$ 
W_{n_1 n_2}^m (\pm k_1, \pm k_2)=\frac{(-1)^m n!}{\Gamma(1\pm k_2)}
$$ 
$$
\times \sqrt{(2 m\pm k_1\pm k_2+1)
\frac{\Gamma(m\pm k_1\pm k_2+1)\Gamma(m\pm k_2+1)}
{n_1! n_2! m! n_\rho!\Gamma(m\pm k_1+1)}} 
$$ 
\begin{eqnarray}
\label{CP16}
\times \sqrt{\frac{\Gamma(n_1\pm k_1+1)\Gamma(n_2\pm k_2+1)}
{\Gamma(n+m\pm k_1\pm k_2+2)}}
{}_3 F_2\biggl(\matrix{-m, \mbox{  }  m\pm k_1\pm k_2+1, \mbox{  }-n_2\hfill\cr
1\pm k_2, \mbox{  }   -n_1-n_2\hfill\cr}
\bigg\vert 1 \biggr)
\end{eqnarray}
Alternatively, by using the formula~\cite{SUS} connecting the Hahn polynomial
$h_n^{(\alpha,\beta)}$ and the function    $_3F_2(1)$, we obtain
$$
W_{n_1 n_2}^m (\pm k_1, \pm k_2)=(-1)^m
\sqrt{(2 m\pm k_1\pm k_2+1)
\frac{m! n_\rho!\Gamma(m\pm k_1\pm k_2+1)}
{n_1! n_2!\Gamma(m\pm k_1+1)\Gamma(m\pm k_2+1)}}
$$
$$
\times \sqrt{\frac{\Gamma(n_1\pm k_1+1)\Gamma(n_2\pm k_2+1)}
{\Gamma(n+m\pm k_1\pm k_2+2)}}
h_m^{(\pm k_2,\pm k_1)}(n_1,n_1+n_2+1)          
$$
in terms of Hahn polynomials.

\subsection{Invariance algebra}

Let us consider the following realization of the SU(1,1) generators 
$$ 
J_0^{(a)} = \frac{1}{4\Omega} \left(
- {\partial}_{x_a x_a}
+ \Omega^2 x_a^2 + \frac{k_a^2 - \frac{1}{4}}{x_a^2}
\right), \; 
J_1^{(a)} = - J_0^{(a)} + \frac{1}{2} \Omega x_a^2, \; 
J_2^{(a)} = \frac{i}{2} \left(
x_a {\partial}_{x_a} + \frac{1}{2} \right) 
$$ 
We thus have two copies (for $a = 1,2$) of the Lie algebra SU(1,1) given by
$$ 
[J_0^{(a)},J_1^{(a)}]= i J_2^{(a)}, \quad 
[J_1^{(a)},J_2^{(a)}]=-i J_0^{(a)}, \quad
[J_2^{(a)},J_0^{(a)}]= i J_1^{(a)}  
$$ 
with the Casimir operator 
\begin{eqnarray}
\label{SY5}
Q_a =  [J_0^{(a)}]^2
      -[J_1^{(a)}]^2 
      -[J_2^{(a)}]^2 
    = \frac {1}{4} (k_a^2 - 1)
\end{eqnarray}
Introducing the raising and lowering operators
$J_\pm^{(a)}=J_{1}^{(a)}\pm i J_{2}^{(a)}$, we get 
$$ 
[J_0^{(a)},J_{\pm}^{(a)}] = \pm J_{\pm}^{(a)}, \quad
[J_-^{(a)},J_+    ^{(a)}] =   2 J_    0^{(a)}  \quad \mbox{and} \quad 
Q_a = [J_0^{(a)}]^2 - J_0^{(a)} - J_+^{(a)}  J_-^{(a)}
$$ 
As an irreducible representation of SU(1,1), 
the positive discrete series consists of an infinite number of
states. Each of these states will be
denoted as $|j_am_a\rangle$, where $m_a=j_a+n_a$ ($n_a=0,1,2,\cdots$).
The eigenvalue of the Casimir operator is 
$$ 
Q_a = j_a(j_a-1)
$$ 
so that from (\ref{SY5}) we have $j_a = \frac{1}{2}(1\pm k_a)$. 
The matrix elements of the generators of the group SU(1,1) may 
be obtained through
\begin{eqnarray}
\label{SY8}
J_0^{(a)}|j_a m_a\rangle = m_a|j_a m_a\rangle, \quad
J_{\pm}^{(a)}|j_a m_a\rangle = 
\sqrt{(m_a {\pm} j_a)(m_a {\mp} j_a {\pm} 1)}|j_a m_a {\pm} 1\rangle 
\end{eqnarray}
with $J_-^{(a)}|j_a j_a\rangle = 0$. Let us now define 
\begin{eqnarray}
\label{SY9}
C_0   = J_0  ^{(1)} + J_0  ^{(2)}, \quad C_\pm = J_\pm^{(1)} + J_\pm^{(2)}
\end{eqnarray}
Equation (\ref{SY9}) corresponds to the direct sum of the two SU(1,1) 
algebras for $a=1,2$. The  coupled  basis  $ |j m\rangle $ satisfies 
$$ 
C_0 | j m\rangle = m       | j m\rangle = (j + n)| j m\rangle, \quad 
  Q | j m\rangle = j( j-1) | j m\rangle 
$$ 
Given the values $j_1$ and $j_2$, the parameter 
$j$ can take the discrete values 
$$ 
 j=j_1+j_2+q,  \quad q \in {\bf N} 
$$ 
The Clebsch-Gordan decomposition yields 
$$ 
| j    m   \rangle = \sum_{m_1 m_2} \langle j_1 j_2 m_1 m_2 | j m \rangle \ 
| j_1  m_1 \rangle \otimes
| j_2  m_2 \rangle, \quad  m = m_1 + m_2
$$ 
with
$2 j_a = {1 \pm k_a}$, 
$2 m_a = 2 n_a + 1            \pm k_a$ and 
$2   j = 2q + 2               \pm k_1 \pm k_2$. 
By using the connection between the SU(1,1) CGc and the $_3 F_2(1)$ 
function~\cite{KP}, one can obtain the same hypergeometric function 
as in (\ref{CP16}).

Note that the Hamiltonian $H$ of our two-dimensional oscillator system is
$$ 
H = 2 \Omega C_0
$$ 
From (\ref{SY8}) and (\ref{SY9}), we recover the spectrum of the system 
as given by (\ref{P12}) with $n = n_1+n_2$. 

Let us consider the two following operators
$$ 
N = J_0^{(1)} - J_0^{(2)}, \quad 
M = Q_1 + Q_2 + 2 J_0^{(1)}J_0^{(2)} - J_+^{(1)}J_-^{(2)}
                                     - J_-^{(1)}J_+^{(2)} + \frac{1}{4} 
$$ 
They commute with $H$. Indeed, they are nothing but 
the integrals of motion (\ref{P200}) and (\ref{P22}).
Moreover, let us define a third operator $T$ via $T= [N,M]$. We have 
$$
T = 2 \left[ J_-^{(1)}J_+^{(2)} - J_+^{(1)}J_-^{(2)} \right] 
$$
or
$$ 
T = - \frac{1}{4\Omega}(D_{xx}-D_{yy}) - \frac{i}{2\Omega}D_{xy}L_z 
+ 
\frac{k_1^2-\frac{1}{4}}{2\Omega x^2}\left( y{\partial}_y + \frac{1}{2} \right) 
-
\frac{k_2^2-\frac{1}{4}}{2\Omega y^2}\left( x{\partial}_x + \frac{1}{2} \right)
$$ 
The operators $N$, $M$, $T$ and $H$ span a closed quadratic algebra
since
 $$ 
 [M , T] = -2 (MN + NM) + \frac{k_1^2-k_2^2}{2\Omega} H-N, \; 
 [T , N] = -2  N^2+\frac{1}{2\Omega^2} H^2-4 M-k_1^2-k_2^2-1
 $$ 
hold in addition to $[N,M] = T$, $[N,H]=0$ and $[M,H]=0$. 

In the limiting case $k_1=k_2=\frac{1}{2}$, we obtain a quadratic algebra
too. In this case 
$$ 
N=   \frac{1}{4\Omega}(D_{xx}-D_{yy}), \quad M=\frac{1}{4}L_z^2, \quad
T= - \frac{1}{4\Omega}(D_{xx}-D_{yy}) - \frac{i}{2\Omega}D_{xy}L_z 
$$ 
Instead of $N$, $L_z^2$ and $T$, we can consider $N$, $L_z$ and $[N,L_z]$. 
In this regard, by putting
$$ 
P_1 = N, \quad P_2 = \frac{1}{2}L_z, \quad P_3 = \frac{1}{i}[P_1,P_2] 
                                               = \frac{1}{2} \Omega D_{xy}
$$ 
we end up with the Lie algebra corresponding to the commutation relations
$$ 
[P_k,P_{\ell}]= i \varepsilon_{k \ell m} P_m, \quad k,\ell,m \in \{ 1,2,3 \}
$$ 

Finally, going back to the generic case for $k_1$ and $k_2$, we define 
$$ 
L_0 = - \frac{1}{4}( 2 N \mp k_1 \pm k_2 ) 
$$
and
$$ 
L_+ =   \frac{J_-^{(1)} J_+^{(2)}} {\sqrt{(n_1 \pm k_1)(n_2 \pm k_2 + 1)}},
\quad
L_- =   \frac{J_+^{(1)} J_-^{(2)}} {\sqrt{(n_2 \pm k_2)(n_1 \pm k_1 + 1)}}
$$ 
They act on the eigenfunctions (\ref{P1}) of the Hamiltonian $H$ as 
$$ 
L_0 \Psi_{n_1 n_2} = \frac{1}{2} (n_2-n_1)     \Psi_{n_1 n_2},
\quad
L_{\pm} \Psi_{n_1 n_2} = \sqrt{\left(n_1 + \frac{1}{2} \mp \frac{1}{2}\right)
                               \left(n_2 + \frac{1}{2} \pm \frac{1}{2}\right)}
        \Psi_{n_1 \mp 1 n_2 \pm 1} 
$$ 
The operators $L_0$, $L_+$ and $L_-$ generate 
the Lie algebra SU(2) with 
$$ 
[L_0,L_{\pm}] = \pm L_{\pm}, \quad
[L_+,L_-    ] =   2 L_0
$$ 
and are closely connected to our integrals of motion.

\section{Three-dimensional case}

\subsection{Spherical basis}

In spherical coordinates ($r, \theta, \varphi$), 
the potential (\ref{I1}) can be rewritten as
$$ 
V = \frac{1}{2    } \Omega^2 r^2 +
    \frac{1}{2 r^2}
\left[
\frac{1}{\sin^2\theta}
\left(
\frac{ k_1^2 - \frac{1}{4} }{\cos^2\varphi} +
\frac{ k_2^2 - \frac{1}{4} }{\sin^2\varphi}
\right) +
\frac{ k_3^2 - \frac{1}{4} }{\cos^2\theta }
\right]
$$ 
Looking for a solution of Eq.~(\ref{SCH}) in the form 
$R(r)\Theta(\theta)\Phi(\varphi)$, we are left with the system
\begin{eqnarray}
\label{S4}
\left( d_{\varphi \varphi} 
+ A^2 - \frac{k_1^2 - \frac{1}{4}}{\cos^2\varphi}
- \frac{k_2^2 - \frac{1}{4}}{\sin^2\varphi} \right) \Phi
&=& 0
\\[2mm]
\label{S5}
\left[ 
\frac{1}{ \sin \theta } d_{\theta}( \sin \theta d_{\theta} )
+ J(J+1)-\frac{A^2}{\sin^2\theta}-
\frac{k_3^2 - \frac{1}{4}}{\cos^2\theta} \right] \Theta &=& 0 
\\[2mm]
\label{S6}
\left[
\frac{1}{r^2} d_{r} ( r^2 d_{r} ) 
+ 2E - \Omega^2 r^2 - \frac{J(J+1)}{r^2} \right] R  &=& 0 
\end{eqnarray}

The solution 
$ \Phi(\varphi) \equiv \Phi_m(\varphi ; \pm k_1, \pm k_2) $
of Eq.~(\ref{S4}), satisfying  the  boundary conditions (\ref{P8}) 
and the normalization condition (\ref{aadd}), 
is given by (\ref{P9}). The separation constant
$A$ in (\ref{S4}) and (\ref{S5}) is quantized 
according to (\ref{P10}).

The solution 
$\Theta (\theta) \equiv \Theta_{qm} (\theta ; \pm k_1 \pm k_2 \pm k_3)$ 
of (\ref{S5}) is (see~\cite{E1})
\begin{eqnarray}
\Theta(\theta)&=&
\sqrt{\frac{[2(m+q+1) \pm k_1 \pm k_2 \pm k_3]
q!\Gamma(q +2 m \pm k_1 \pm k_2 \pm k_3 +2)}
{\Gamma(q \pm k_3+1)\Gamma(q +2 m +2 \pm k_1 \pm k_2 )}} \nonumber
\\[2mm]
& \times &(\cos\theta)^{\frac{1}{2} \pm k_3}
(\sin \theta)^{A} P_{q}^{(A ,\pm k_3)}(\cos 2\theta)     \nonumber
\end{eqnarray}
which satisfies the boundary condition
$$ 
\Theta(0)=\Theta(\frac{\pi}{2})=0
$$ 
and the normalization condition
$$ 
2 \int_{0}^{\frac{\pi}{2}}
  \Theta_{q' m}(\theta ; \pm k_1, \pm k_2, \pm k_3)^*
  \Theta_{q  m}(\theta ; \pm k_1, \pm k_2, \pm k_3)   \sin \theta d{\theta}=
  \delta_{q'{q}}
$$ 
The spherical separation constant $J$ in (\ref{S5}) 
and (\ref{S6}) is 
$$ 
J = 2 q+A\pm k_3+\frac{1}{2} = 2q + 2m \pm k_1 \pm k_2 \pm k_3+\frac{3}{2}
$$ 

The solution 
$R(r) \equiv R_{n_r     q m} (r ; \pm k_1, \pm k_2, \pm k_3)$
of Eq.~(\ref{S6}) is
$$ 
R(r) =
\sqrt{\frac{2{\Omega}^{\frac{3}{2}} n_r!}
{\Gamma(n_r + 2q + 2m \pm  k_1 \pm k_2 \pm k_3+3)}} 
\left( \sqrt{\Omega r^2} \right)^{J}
{\rm exp}\left(-\frac{\Omega}{2}r^2\right)
L_{n_r}^{J+\frac{1}{2}}(\Omega r^2)
$$ 
with
$$ 
\int_{0}^{\infty}
R_{n_r^{'} q m} (r ; \pm k_1, \pm k_2, \pm k_3)
R_{n_r     q m} (r ; \pm k_1, \pm k_2, \pm k_3) r^2 d r =
\delta_{n_r^{'} n_r}
$$ 
where $n_r \in {\rm {\bf N}}$ is the radial quantum number.

The energy of the system is
$$ 
E =\Omega(2 n_r+J+\frac{3}{2}) =
\Omega (2n  \pm k_1 \pm k_2 \pm k_3+ 3), \quad
n \in {\bf N} 
$$ 
where $n = n_r + q + m$ is the principal quantum number. It corresponds 
to the wavefunctions
$$
\Psi_{n_r q m } (r , \theta , \varphi ; \pm k_1 , \pm k_2 , \pm k_3)
\equiv R (r) \Theta (\theta) \Phi (\varphi)
$$
with $n$ fixed. 

\subsection{Cylindrical basis}

In cylindrical coordinates ($\rho, \varphi, z$), we have
$$ 
V = \frac{1}{2} \Omega^2 \rho^2  +
\frac{1}{2 \rho^2}\left(\frac{k_1^2 - \frac{1}{4}}{\cos^2\varphi} +
\frac{k_2^2 - \frac{1}{4}}{\sin^2\varphi}\right) +
\frac{1}{2}\left(\Omega^2 z^2+ \frac{{k_3}^2 - \frac{1}{4}}{z^2}
\right)
$$ 
The corresponding Schr\"odinger equation may be solved 
by looking for a solution in the form 
$R (\rho) \Phi (\varphi) Z (z)$. 
By combining the results of Sections 2 and 3, we get 
$$
Z(z) \equiv \Psi_{n_3}(z ; \pm k_3), \quad
\Phi(\varphi) \equiv \Phi_{m}(\varphi ; \pm k_1 , \pm k_2), \quad
R(\rho) \equiv R_{n_{\rho} m} (\rho ; \pm k_1 , \pm k_2)
$$
as given by (\ref{C4}), (\ref{P9}) and (\ref{P11}), respectively. The energy 
$$ 
E = \Omega(2n \pm k_1 \pm k_2 \pm k_3 + 3)
$$ 
corresponds to the wavefunctions 
$$
\Psi_{n_{\rho} m n_3} (\rho, \varphi, z ; \pm k_1, \pm k_2, \pm k_3) 
\equiv R (\rho) \Phi (\varphi) Z(z)
$$
for which the principal quantum number $n = n_\rho + m + n_3$ is fixed.

\subsection{Connecting Cartesian, cylindrical and spherical bases}

In the three-dimensional case, we have
$$ 
{\Psi}_{n_1 n_2 n_3} =
\sum_{m = 0}^{n_1 + n_2} W_{n_1 n_2}^{m} (\pm k_1,\pm k_2 )
 \Psi_{n_\rho m n_3}, \; 
{\Psi}_{n_\rho m n_3} =
\sum_{q = 0}^{n_\rho + n_3} V_{n_\rho n_3}^{q}
(\pm k_1, \pm k_2, \pm k_3)
\Psi_{n_r q m}  
$$ 
where $n_1 + n_2 =m + n_\rho$ and
      $n_r + q =n_\rho + n_3$. For the expansion of the Cartesian
basis over the spherical basis, we have
 \begin{eqnarray}
 \label{CS3}
 {\Psi}_{n_1 n_2  n_3} =
 \sum_{mq} C_{n_1 n_2 n_3}^{m q} (\pm k_1, \pm k_2, \pm k_3)
 \Psi_{n_r q m} 
 \end{eqnarray}
where $n_1 + n_2 + n_3 = n_r + q + m$.
The coefficient $W_{n_1 n_2}^{m} (\pm k_1, \pm k_2)$ is identical 
to the one found in the two-dimensional case. It is given by 
(\ref{CP10}). Similarly, it is easy to obtain
\begin{equation}
\label{CS5}
V_{n_\rho n_3}^{q} (\pm k_1, \pm k_2, \pm k_3) = (-1)^{n_\rho-q}
{} \langle a'b' \alpha' \beta'|c' \gamma' \rangle 
\end{equation}
where
$2 a' = {n_3+n_\rho \pm k_3}$,
$2 b' = {n_3+n_\rho+2 m +1 \pm k_1 \pm k_2}$,
$2 c' = 2 q + 2 m  + {1 \pm k_1 \pm k_2 \pm k_3}$, 
$2 \alpha'= {n_3-n_\rho \pm k_3}$ and 
$2 \beta' = 2 m+{n_\rho - n_3 +1 \pm k_1 \pm k_2}$. 
The expansion coefficients in (\ref{CS3}) are given by the formula 
\begin{equation}
\label{CS6}
C_{n_1 n_2 n_3}^{m q} (\pm k_1, \pm k_2, \pm k_3) = 
W_{n_1    n_2}^{m} (\pm k_1, \pm k_2)
V_{n_\rho n_3}^{q} (\pm k_1, \pm k_2, \pm k_3) 
\end{equation}
The value of the right-hand side of (\ref{CS6}) follows from 
(\ref{CP10}) and (\ref{CS5}).

\section{Acknowledgments}

The authors would like to thank V.M. Ter-Antonyan for interesting discussions.

\end{document}